\documentclass[12pt]{article}
\usepackage{graphicx}

\begin{document}

Astrophysics, Vol. 54, No. 3, 2011\\

\begin{center}
{\bf \Large SURVEY OF $H\alpha$ EMISSION FROM THIRTY NEARBY DWARF GALAXIES}\\

\bigskip

{\bf S. S. Kaisin, I. D. Karachentsev, and E. I. Kaisina}

Special Astrophysical Observatory, Russian Academy of Sciences, Russia;\\ 
e-mail: skai@sao.ru

\end{center}

{\em Measurements of the $H\alpha$ flux from 30 neighboring dwarf galaxies are presented. After correction for absorption, these fluxes are used to estimate the star formation rate (SFR). The SFR for 18 of the galaxies according to the $H\alpha$ emission are compared with estimates of the SFR from FUV magnitudes obtained with the GALEX telescope. These are in good agreement over the range $\log[SFR] = [-3,0]M_{\odot}$/year.}

\noindent Keywords: {\em dwarf galaxies, star formation}

\bigskip

{\bf 1. Introduction} \\

\bigskip

This article is a continuation of a cycle of papers [1--6] on a systematic survey in the $H\alpha$
 line of all the galaxies in the Local volume with distances $D < 10$ Mpc. The goal of this project is to measure the$H\alpha$  fluxes and determine the star formation rate (SFR) of all the galaxies, regardless of their luminosities, morphological types, densities of their environment, and other parameters. As a result, we plan to obtain a standard sample that is free of selectivity effects for analyzing the conditions under which  gas into stars is transformed.

 Various data on 450 galaxies in the Local volume have been collected in the ``Catalog of Neighboring Galaxies'' (CNG) [7], which, however, contains no data on their $H\alpha$ fluxes. Over the past few years observations of about 250 nearby galaxies have been made with the 6-m BTA telescope. In combination with other $H\alpha$ surveys [8--10], this led to almost complete coverage of the sample of northern galaxies in the Local volume. In the meantime, systematic surveys of the sky (SDSS, 2dF, 6dF) in the optical range and in the HI neutral hydrogen line (HIPASS, ALFALFA) have increased the number of galaxies at distances within 10 Mpc from 450 to 800. Thus, we have had
to continue our  $H\alpha$ survey program. Here we present data on thirty nearby galaxies, most of which are shown in the $H\alpha$ line for the first time.

\bigskip

{\bf 2. Observations and data processing}

\bigskip

The nearby dwarf galaxies were observed using the 6-m telescope at the Special Astrophysical Observatory of the Russian Academy of Sciences (SAO RAN) during 2009--2011 with the SCORPIO reducer [11]. Images of the galaxies were recorded with a $2048\times2048$ pixel CCD which provided a 6.1$^{\prime}$ field with a resolution of 0.18$^{\prime\prime}$/pixel. Images of the galaxies in the $H\alpha$ line and the neighboring continuum were obtained using a narrow band $H\alpha$ interference filter with a width $\Delta\lambda = 74$\AA \ and an effective wavelength $\lambda_{eff} = 6555$\AA, along with two medium band filters, SED 607 with  $\Delta\lambda = 167$\AA \ and $\lambda_{eff} = 6063$\AA \ and SED 707 with $\lambda_{eff} = 207$\AA \ and $\lambda_{eff} = 7063$\AA. The typical exposure time was 2$\times$600 s in the $H\alpha$ line and 2$\times$300 s in the continuum. Because of the small range of radial velocities in the nearby galaxies, the $H\alpha$ images of all of them could be taken with the same filter.
\begin{table}[b]
\caption{$H\alpha$ Fluxes from Nearby Galaxies}
\begin{tabular}{lcrccccrrr} \\ \hline

Name   &    RA,    Dec   & T & $V_{LG}$ & 
$D_{MW}$  & $M_B$ & $\log M_{HI}$ &  $\log F_{H\alpha}$  &$\log F_{H_\alpha}$ & $\log SFR$\\
	  &  (2000.0)   &    & km s$^{-1}$   &Mpc  & mag  & $M_{\odot}$  &   obs   & corr&  $M_{\odot}$/y \\
\hline
U288    & 002904.0+432554 &10    & 464&  6.7\, & $-$13.82&  7.73 &  $-$13.33 &$-$13.26&  $-$2.51\\
U685    & 010722.3+164102 & 9    & 349&  4.51& $-$14.31&  7.68 &  $-$12.64& $-$12.59 & $-$2.18\\
U731    & 011044.0+493608 & 7    & 901& 12.5 \,h& $-$16.09&  8.97 &  $-$12.79 &$-$12.66 & $-$1.36\\
U1171   & 013941.6+155411 &10    & 906&  7.3 & $-$13.86&  7.25 &  $-$14.64& $-$14.59 & $-$3.76\\
DDO13   & 014009.6+155417& 10    & 798&  9.0 & $-$15.72&  8.73 &  $-$12.66& $-$12.60 & $-$1.59\\
KDG10   & 014341.4+154123 &10    & 953&  7.9 & $-$13.53&  7.72&   $-$13.47& $-$13.40 & $-$2.50\\
d0224+41& 022420.7+410212 &10    & ---& 10.5 \,m& $-$12.63&  $<$7.4& $<-$15.4& $<-$15.4&  $<-$4.3\\
d0226+33 &022652.8+332537 &10    & 709&  9.3 \,m& $-$12.96&  7.0 &   $<-$15.6& $<-$15.6&  $<-$4.6\\
d0243+37 &024302.0+375926 &$-$3  & --- &  10.5 \,m& $-$13.34&  $<$7.4& $<-$15.3& $<-$15.3 & $<-$4.2\\
N1156   & 025942.4+251415 & 9    & 510&  7.8 & $-$17.70&  8.93 &  $-$11.46 &$-$11.25&  $-$0.37\\
U2684   & 032023.7+171742 &10    & 438&  6.5 & $-$13.37&  8.00&   $-$13.57 &$-$13.44 & $-$2.71\\
U2716   & 032407.2+174515 & 8    & 467&  6.5 \,m& $-$15.01&  7.84 &  $-$12.98 &$-$12.85 & $-$2.12\\
N2337    &071013.6+442725  &9    & 477&  7.9 & $-$16.51&  8.77 &  $-$11.86 &$-$11.78&  $-$0.88\\
U3817   & 072244.5+450630 &10    & 478&  8.6 & $-$14.15&  8.24 &  $-$13.13& $-$13.04 & $-$2.07\\
KKH46   & 090836.6+051732 &10    & 409&  5.7 \,h& $-$11.93&  7.31 &  $-$13.28 &$-$13.24 & $-$2.63\\
KDG56   & 093012.9+195930 &10    & 440&  8.9 \,m& $-$12.92&  7.17 &  $-$15.34 &$-$15.30 & $-$4.30\\
U5086    &093248.9+212754 &$-$3  & 394&  8.9 \,m& $-$13.95& $<$5.75& $<-$15.5 &$<-$15.5 & $<-$4.5\\
KDG58   & 094027.0 +000333& 10 & --- &10.0: &$-$13.05& --- & $-$15.37 &      $-$15.32&      $-$4.22\\
U5186    &094259.8+331552 &10    & 500&  6.9 \,h& $-$12.98&  7.23 &  $-$14.84& $-$14.83 & $-$4.05\\
U5288   & 095117.2+074938 &10    & 377&  6.8 & $-$14.74&  8.44  & $-$12.47 &$-$12.44 & $-$1.67\\
KKH60    &101559.4+064821& 10    &1188& 16.5 \,h& $-$13.68&  7.88 &  $<-$14.5 &$<-$14.5 & $<-$3.0\\
N4523   & 123348.0+151005 & 8    & 173& 17.0 \,m& $-$17.03&  9.14 &  $-$12.11 &$-$12.07 & $-$0.51\\
U7857   & 124154.2+134622&$-$1   &$-$73&17.0 \,m& $-$16.57&  $<$8.0& $<-$15.0 &$<-$15.0 & $<$-3.4\\
U8061   & 125643.4+115552 &10    &471&   6.5 \,h& $-$13.99&  7.30 &  $-$13.65 &$-$13.62 & $-$2.89\\
U8245   & 130834.2+785613 & 8    &273&   3.7 \,h& $-$13.49&   ---  & $-$13.38 &$-$13.35 & $-$3.11\\
KKH86    &135433.6+041435 &10    &209&   2.60& $-$10.30&  5.90  & $<-$15.9  &$<-$15.9  & $<-$6.0 \\
N6503 & 174927.6 +700841 & 6&301&5.27& $-$18.08& 9.01& $-$11.07& $-$11.04& $-$0.50\\
U11411   &190842.3+701702 & 9    &345&   4.7 \,h& $-$13.89&   ---  & $-$12.45 &$-$12.34 & $-$1.90\\
N6789    &191641.8+635822 &10    &144&   3.60& $-$14.32&  6.5  &  $-$12.75 &$-$12.69 & $-$2.48\\
KKH98    &234534.0+384304 &10    &151&   2.45& $-$10.78&  6.76  & $-$13.49 &$-$13.38 & $-$3.50\\
\hline
\end{tabular}
\end{table} 
The observational data were processed in the standard way using the MIDAS package. After the electronic bias frame was subtracted, the image was normalized to a flat field obtained at twilight. Then traces of cosmic particles were eliminated by comparing images and subtracting the continuum normalized from the images of 5--15 stars in the field of the frame. The measured integral $H\alpha$ fluxes of the galaxies were calibrated using images of stars from the spectrophotometric standards of Oke [12] obtained on the same night. The internal errors for the $H\alpha$ flux measurements were typically about 15\%
. The major contribution to the error was from changes in the atmospheric conditions. The resultant $H\alpha$ fluxes also contained a contribution from the [NII] emission doublet, which is small in the case of dwarf galaxies, so we neglected it. 

Images of the galaxies we have observed are shown as a mosaic in Fig. 1. The left and right images of each galaxy correspond to the sum and difference of images taken in $H\alpha$ and in the continuum. The angular size of the images is $4^{\prime}\times4^{\prime}$ and north and east directions are indicated by arrows. The major characteristics of these galaxies are listed in Table 1, the columns of which give the following: (1) galaxy name; (2) equatorial coordinates at epoch 2000.0; (3) morphological type on the Vaucouleurs digital scale; (4) radial velocity of the galaxy relative to the centroid of the Local group; (5) distance to the galaxy from the CNG catalog [Here the values with two significant digits were determined using the tip of the red giant branch and those with one significant digit, using the luminosity of the brightest stars. Some distances were determined from the membership of a galaxy in a group or cluster (``{\em m}'') or from the radial velocity ``{\em h}'' for the Hubble constant $H_0 = 72$ km s$^{-1}$Mpc$^{-1}$]; (6) absolute magnitude of the galaxy corrected for absorption of light in accordance with  [13]; (7) the logarithm of the hydrogen mass of the galaxy from the Lyon Extragalactic Database (LEDA); (8, 9) the observed and extinction-corrected $H\alpha$ flux of the galaxy on a logarithmic scale in units of erg/cm$^2$/s; and, (10) integral star formation rate in the galaxy, $[SFR] = 1.27\cdot 10^9\cdot F_c (H\alpha)\cdot D^2$ , where the distance to it is expressed in Mpc [14].

\bigskip

{\bf 3. Some features of the observed galaxies}

\bigskip 

UGC 288. An isolated dIrr galaxy, the distance to which has been estimated from the luminosity of the brightest stars [7]. A star of medium brightness is projected onto its center.

 UGC 685 = KIG 45. A BCD galaxy with active star formation in its center. It is listed in the catalog of isolated galaxies [15]. The distance to it was determined from the tip of the red giant branch. 

UGC 731 = DDO 9. This is an Sd galaxy, the distance to which has been estimated from the radial velocity with a Hubble constant $H_0 = 72$ km s$^{-1}$Mpc$^{-1}$. Small emission knots are distributed asymmetrically over the galactic disk. Detailed $R$-photometry and the velocity field for this galaxy have been obtained by Swaters [16].

 UCG 1171, UGC 1176 = DDO 13, KDG 10. Three dIrr galaxies associated with the bright spiral galaxy NGC 628. The distances to them have been determined from the luminosity of the brightest stars. 

d0224+41, d0226+33, d0243+37. Three dwarf systems in a group around NGC 1023 discovered by
Karachentsev, et al. [17] The southernmost of the three has a very low surface brightness and lies in the remote periphery of the spiral NGC 925. Our observations yield only an upper limit to the $H\alpha$ flux for them.

 NGC 1156=KIG 121. An isolated BCD galaxy, the distance to which was determined from the brightest stars. Strong star formation sites are observed in the center and the NE side of the galaxy. 

UGC 2684, UGC 2716. A wide pair of dwarf galaxies with similar radial velocities. The distance to the first is estimated from the brightest stars. 

UGC 2337. An isolated BCD galaxy with a bright emission bar. The distance to it was determined from the luminosity of the brightest stars. 

UGC 3817. An irregular galaxy with a set of faint emission knots. Probably it and NGC 2337 form a wide pair with a difference of radial velocities of just 1 km s$^{-1}$. 

KKH 46. An isolated irregular galaxy in which five emission knots form an equilateral triangle. 

KDG 58 = KKH 53. An irregular galaxy with a very low surface brightness and without signs of HI emission in the HIPASS survey. The distance to it is nominally 10 Mpc. A faint HII region, indicated in the figure by a circle, can be seen at the optical boundaries of this galaxy. 

KDG 56=KKH 51, UGC 5086. These dIrr and dSph systems are dwarf satellites of the bright spiral NGC 2903. No signs of star formation can be seen in these two galaxies. 

UGC 5186, UGC 5288. Isolated dIrr galaxies, the distances to which have been determined, respectively, from the radial velocity and from the brightest stars. 

KKH 60. This dIrr galaxy has been assigned to the Local volume based on a radial velocity $V_{LG}$ = +108 km/ s measured in a noisy optical spectrum [18]. However, its radial velocity $V_{LG}$ = +1188 km s$^{-1}$ in the HI line [19] indicates that the galaxy lies outside the Local volume at a distance of 16.5 Mpc. A shift of the $H\alpha$ line by 30\AA \ leads to an underestimate in our measured $H\alpha$ flux.

NGC 4523, UGC 7857. Two galaxies of types Sdm and S0/dE with low radial velocities. These are not counted as among the Local volume, since they probably belong to the Virgo cluster.

UGC 8061. An isolated dIrr galaxy projected onto the eastern edge of the Virgo cluster. The distance to it, determined from the radial velocity, requires independent confirmation. 

UGC 8245. A dwarf galaxy at a high declination. Its optical radial velocity $V_h = +70\pm59$, km s$^{-1}$ has been measured by Falco et al. [20]. Its heliocentric radial velocity is $+58\pm55$ km s$^{-1}$ according to [21].

 KKH 86. A nearby dIrr galaxy without visible signs of star formation. The distance to it is 2.60 Mpc, measured from the tip of the red giant branch. 

NGC 6503 = KIG 837. An isolated Scd galaxy at a distance of 5.27 Mpc measured from the tip of the red giant branch. 

UGC 11411. A dwarf BCD galaxy at a high declination. Its low optical radial velocity $V_h = +69\pm38$ km/ s measured by Falco et al. [20], was confirmed by our observations with the 6-m telescope, which gave $V_h = +45\pm35$ km s$^{-1}$.
 UGC 11411 may be a distant satellite of NGC 6503 at a projected distance of $\sim$500 kpc. Three bright compact HII regions stand out in the body of UGC 11411, embedded in a common diffuse emission medium. UGC 11411 has one of the highest specific star formation rates per unit luminosity in the Local volume.

 NGC 6789=KK 245. A nearby BCD galaxy at a distance of 3.60 Mpc measured from the tip of the red giant branch. Its $H\alpha$ emission is concentrated in the central region.

 KKH 98. A nearby isolated dIrr galaxy with a faint diffuse $H\alpha$ emission over the entire disk. The distance to it, 2.45 Mpc, was determined from the tip of the red giant branch.

\bigskip

{\bf 4. Comparison of the $H\alpha$ fluxes with other data}

\bigskip

The accuracy of determining the $H\alpha$ flux from a galaxy is affected by a number of factors, such as the stability of the atmospheric conditions during the observations, the accuracy with which the sky background is subtracted from the images in both $H\alpha$ and the continuum, the accounting for the contribution of the distant edge in extended galaxies, differences in the way the absorption of light and the contribution of the [NII] doublet are taken into account, etc. Lee, et al. [22] have shown that the relative contribution of the [NII] doublet for dwarf galaxies is usually less than 10\%
. Errors in background subtraction for galaxies with low surface brightnesses, as well as the possible neglect of a diffuse emission component between compact HII regions, can make a substantial contribution to the uncertainty in the estimated $H\alpha$ flux. 

Of the observed galaxies, there were 11 objects for which the $H\alpha$ fluxes have been measured by others, besides ourselves. These objects are listed in Table 2. The first column gives the name of the galaxy, the second, our estimate of the $H\alpha$ flux, and the third, the fluxes obtained by others, with the corresponding references indicated in the last column. A comparison of our (``KK'') and the other (``lit'') estimates shows that the mean square difference of the logarithm of the fluxes is 0.116, and the average arithmetic difference $\langle\log F_{KK} − \log F\rangle = +0.045\pm0.026$, i.e., our values of the $H\alpha$ flux are 11\%
 higher than the others on the average. This may be partly because the faint diffuse emission between HII regions was not taken into account in all the reports. According to the data in col. (3), the mean square error of the $H\alpha$ flux measurements for the various reports is 0.069 dex. Taking the squared difference of these from the mean square difference of the estimates, 0.116 dex, we find a typical error of 0.093 dex for our measurements, or somewhat higher than our internal estimate ($\sim$0.06 dex or 15\%
).

\begin{table}[hbt]
\caption{Comparison of the $H\alpha$ Fluxes with Published Values}
\begin{tabular}{lccl}\\ \hline

Name&  $\log F_{KK}$  & $\log F_{lit}$   &   Ref.\\
\hline
 U685  &  $-$12.64   &  $-$12.57$\pm$.07  & [10]\\
 U685  &              & $-$12.62$\pm$.01 & [25]\\
 DDO13 &  $-$12.66   &  $-$12.87$\pm$.01  & [9]\\
 DDO13  &             & $-$12.81$\pm$.05 &  [10]\\
 N1156  & $-$11.46    & $-$11.51$\pm$.03  &[10]\\
 N1156  &             & $-$11.56$\pm$.04 &  [8]\\
 N1156  &             & $-$11.63$\pm$.01 & [9]\\
 U2684  & $-$13.57    & $-$13.72$\pm$.10 &  [10]\\
 U2684  &             & $-$13.70$\pm$.20 &  [8]\\
 U2716  & $-$12.98    & $-$12.95$\pm$.03 &  [10]\\
 N2337  & $-$11.86    & $-$11.86$\pm$.03  & [10]\\
 N2337  &             & $-$11.87$\pm$.04 &  [8]\\
 U3817  & $-$13.13    & $-$12.97$\pm$.05 &  [10]\\
 U3817  &             & $-$13.10$\pm$.06 &  [8]\\
 U5288  & $-$12.47    & $-$12.47$\pm$.04  &[10]\\
 U5288  &             & $-$12.53$\pm$.01 &  [25]\\
 U8245  & $-$13.38    & $-$13.46$\pm$.14  & [10]\\
 N6503  & $-$11.07    & $-$11.29$\pm$.07  &[10]\\
 N6789  & $-$12.75    & $-$12.58$\pm$.05 &  [10]\\
 N6789   &            & $-$12.56$\pm$.03  & [24]\\
 N6789  &             & $-$12.89$\pm$.04 &  [9]\\

\hline
\end{tabular}
\end{table}

\bigskip

{\bf 5. Discussion}

\bigskip
Determining the integral star formation rate in a galaxy (SFR) on the basis of its integral $H\alpha$ flux involves a whole series of assumptions regarding the parameters of the initial stellar mass function, properties of scattering of $H\alpha$ photons in the interstellar medium, and the distribution of dust surrounding young blue stars. In recent years there has been a tendency to refine and revise these assumptions [23]. On the average, the star formation rate determined from $F(H\alpha$) is characterized by a time scale of $\sim5-10$ million years. New and independent possibilities for determining the SFR appeared with the launching in 2003 of the GALEX ultraviolet satellite. This 50-cm diameter telescope with a field of view of 1.2$^{\circ}$ and an angular resolution of $\sim5^{\prime\prime}$ has been used to measure the ultraviolet fluxes $F(FUV)$ from several hundred nearby galaxies at an effective wavelength of 1539\AA \ with a width FWHM = 269\AA [24]. The integrated flux $F(FUV)$ from young stars in a galaxy yields a direct estimate of the star formation rate over a time scale of $\sim50-100$ years. However, even this method has a significant shortcoming because of the uncertainty in the amount of absorption by the dust surrounding young stars.

Using the values of the $FUV$ fluxes given in [25], as well as in the NED, we have determined the integrated star formation rate for 19 of the galaxies that we have observed. The estimates of the SFR obtained from the $H\alpha$

\begin{table}
\caption{Star Formation Rates According to $H\alpha$ and $FUV$ Fluxes}
\begin{tabular}{lrr} \\ \hline
Name & $\log SFR_c$& ($M_{\odot}yr^{-1})$ \\
&       $H\alpha$ & $FUV$ \\
\hline
N1156 & $-$0.37& $-$0.48\\
N6503& $-$0.50& $-$0.77\\
N2337& $-$0.88& $-$1.11\\
DDO13 & $-$1.59 & $-$1.48 \\
U5288& $-$1.67& $-$1.83\\
U11411& $-$1.90& $-$2.35\\
U3817& $-$2.07& $-$2.17\\
U2716& $-$2.12& $-$1.97\\
U685 & $-$2.18 & $-$2.32 \\
N6789& $-$2.48& $-$2.59\\
KKH46& $-$2.63& $-$2.68\\
U2684 &$-$2.71& $-$2.26\\
U8245& $-$3.11& $-$2.68\\
KKH98& $-$3.50& $-$3.32\\
U5186& $-$4.05& $-$2.90\\
KDG58& $-$4.22& $-$3.19 \\ 
KDG56& $-$4.30& $-$2.98\\
U5086& $<-$4.5& $<-$4.18\\
KKH86& $<-$6.0& $-$4.17\\
\hline
\end{tabular}
\end{table}

flux and the $FUV$ flux are compared in Table 3, where the galaxies are ordered in terms of their $SFR$. The internal absorption is assumed to be low for most of these dwarf galaxies. Only for two of them, NGC 6503 and NGC 2337, did we introduce corrections $\Delta\log SFR$ for absorption in the $FUV$ band in the amounts of $+0.28$ and $+0.09$, respectively. These data show that, within the range $\log[SFR] = [-3, 0]$, there is complete agreement between the estimates based on $H\alpha$ and $FUV$. The average difference $\log SFR (H\alpha)− \log SFR (FUV)$ is $+0.07\pm0.069$, with the mean square difference of 0.23 dex. For smaller values of $\log[SFR] < -3$ there is a noticeable systematic excess in the SFR values based on the ultraviolet compared to those based on $H\alpha$. The reasons for this difference have been discussed in detail by Lee et al. [22].

 It should be noted that, by now, the $H\alpha$ fluxes have already been measured for $\sim500$ galaxies out of the 800 located in the Local volume ($D < 10$ Mpc). The number of galaxies in this volume with measureable $FUV$ fluxes does not yet exceed 320. Evidently, continuing the ultraviolet survey of nearby dwarf galaxies with an emphasis on objects with especially low luminosities could provide more accurate information on the conditions for star formation in these galaxies when the gas density is extremely low. 

This work was supported by the Russian Foundation for Basic Research (RFFI), grants 10--02--00123 and 10--02--92650.

\newpage

{\bf REFERENCES}

\bigskip

[1]. I. D. Karachentsev, S. S. Kaisin, Z. Tsvetanov, and H. Ford, Astron. Astrophys. 434, 935 (2005). 

[2]. S. S. Kaisin and I. D. Karachentsev, Astrofizika 49, 337 (2006).

[3]. S. S. Kaisin, A. V. Kasparova, A. Yu. Knyazev, and I. D. Karachentsev, Pis’ma v Astron. zh. 33, 1 (2007).

[4]. I. D. Karachentsev and S. S. Kaisin, Astron. J. 133, 1883 (2007). 

[5]. S. S. Kaisin and I. D. Karachentsev, Astron. Astrophys. 479, 603 (2008). 

[6]. I. D. Karachentsev and S. S. Kaisin, Astron. J. 140, 1241 (2010). 

[7]. I. D. Karachentsev, V. E. Karachentseva, W. K. Huchtmeier, and D. I. Makarov, Astron. (=CNG) 

[8]. P. A. James, N. S. Shane, J. E. Beckman, et al., Astron. Astrophys. 414, 23 (2004). 

[9]. D. A. Hunter and B. G. Elmegreen, Astron. J. 128, 2170 (2004). 

[10]. R. C. Kennicutt, J. C. Lee, J. G. Funes, et al., Astrophys. J. Suppl. Ser. 178, 247 (2008). 

[11]. V. L. Afanasiev, E. B. Gazhur, S. R. Zhelenkov, and A. V. Moiseev, Bull. SAO 58, 90 (2005). 

[12]. J. B. Oke, Astron. J. 99, 1621 (1990). 

[13]. D. J. Schlegel, D. P. Finkbeiner, and M. Davis, Astrophys. J. 500, 525 (1998). 

[14]. J. S. Gallagher, D. A. Hunter, and A. V. Tutukov, Astrophys. J. 284, 544 (1984). 

[15]. V. E. Karachentseva, Soobshch. SAO 8, 3 (1973). 

[16]. R. Swaters, Thesis ``Dark Matter in Late-type Dwarf Galaxies,'' Groningen (1999). 

[17]. I. D. Karachentsev, V. E. Karachentseva, and V. K. Huchtmeier, Pis'ma v Astron. zh. 33, 512 (2007).

[18]. D. I. Makarov, I. D. Karachentsev, and A. N. Burenkov, Astron. Astrophys. 405, 951 (2003). 

[19]. M. Haynes (personal communication). 

[20]. E. E. Falco, M. J. Kurtz, M. J. Geller, et al., Publ. Astron. Soc. Pacif. Conf. Ser. 111, 438 (1999). 

[21]. I. D. Karachentsev, E. I. Kaisina, S. S. Kaisin, and L. N. Makarova, Mon. Notic. Roy. Astron. Soc., accepted (astroph/1104. 5318) (2011). 

[22]. J. C. Lee, R. C. Kennicutt, J. G. Funes, et al., Astrophys. J. 692, 1305 (2009). 

[23]. J. Pflamm-Altenburg and P. Kroupa, Astrophys. J. 706, 516 (2009). 

[24]. A. Gil de Paz, B. F. Madore, and O. Pevunova, Astrophys. J. Suppl. Ser. 147, 29 (2003). 

[25]. J. C. Lee, A. Gil de Paz, R. C. Kennicutt, et al., Astrophys. J. Suppl. Ser. 192, 1 (2011). 

[26]. L. van Zee, Astron. J. 119, 2757 (2000). J. 127, 2031 (2004).

\clearpage

\topmargin=-1cm
\begin{figure}
\includegraphics{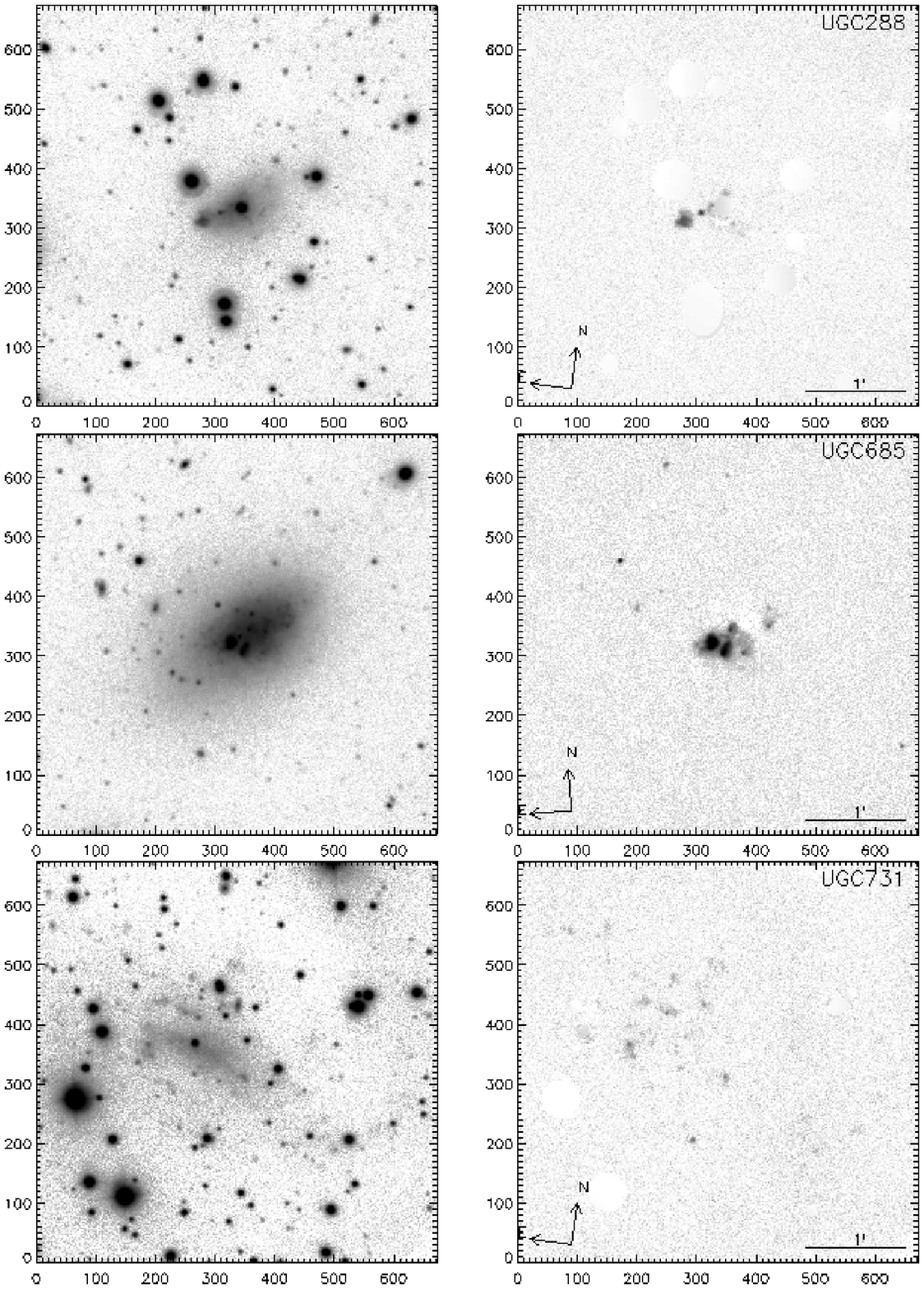}
\caption{Mosaic of images of the 30 nearby dwarf galaxies. On the left are
images in $H\alpha$ +continuum and on the right, in $H\alpha$ with the
continuum subtracted. The size of the pictures is
$4^{\prime}\times4^{\prime}$ and north and east are indicated by arrows.}
\end{figure}

\begin{figure}
\includegraphics{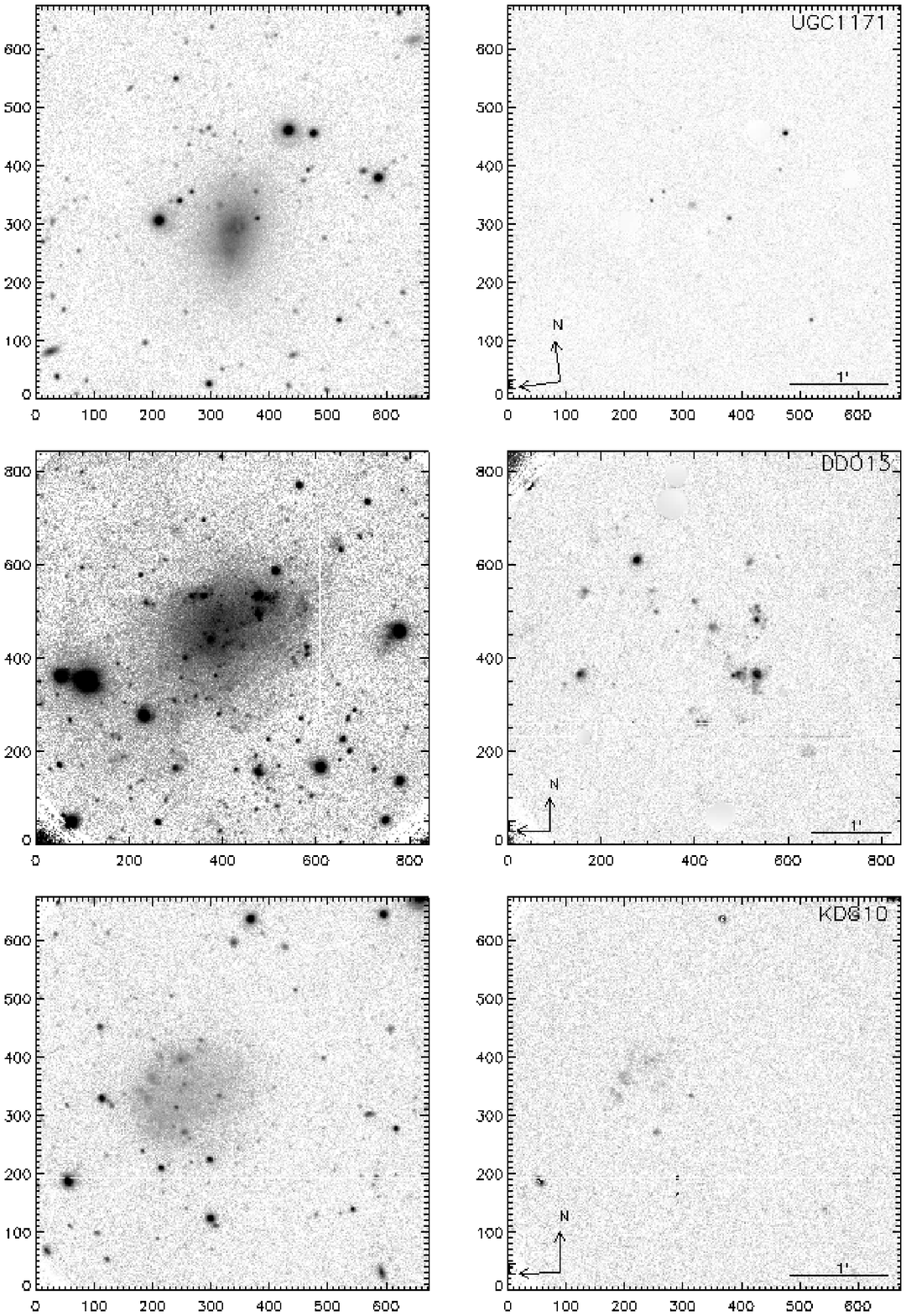}
\end{figure}

\begin{figure}
\includegraphics{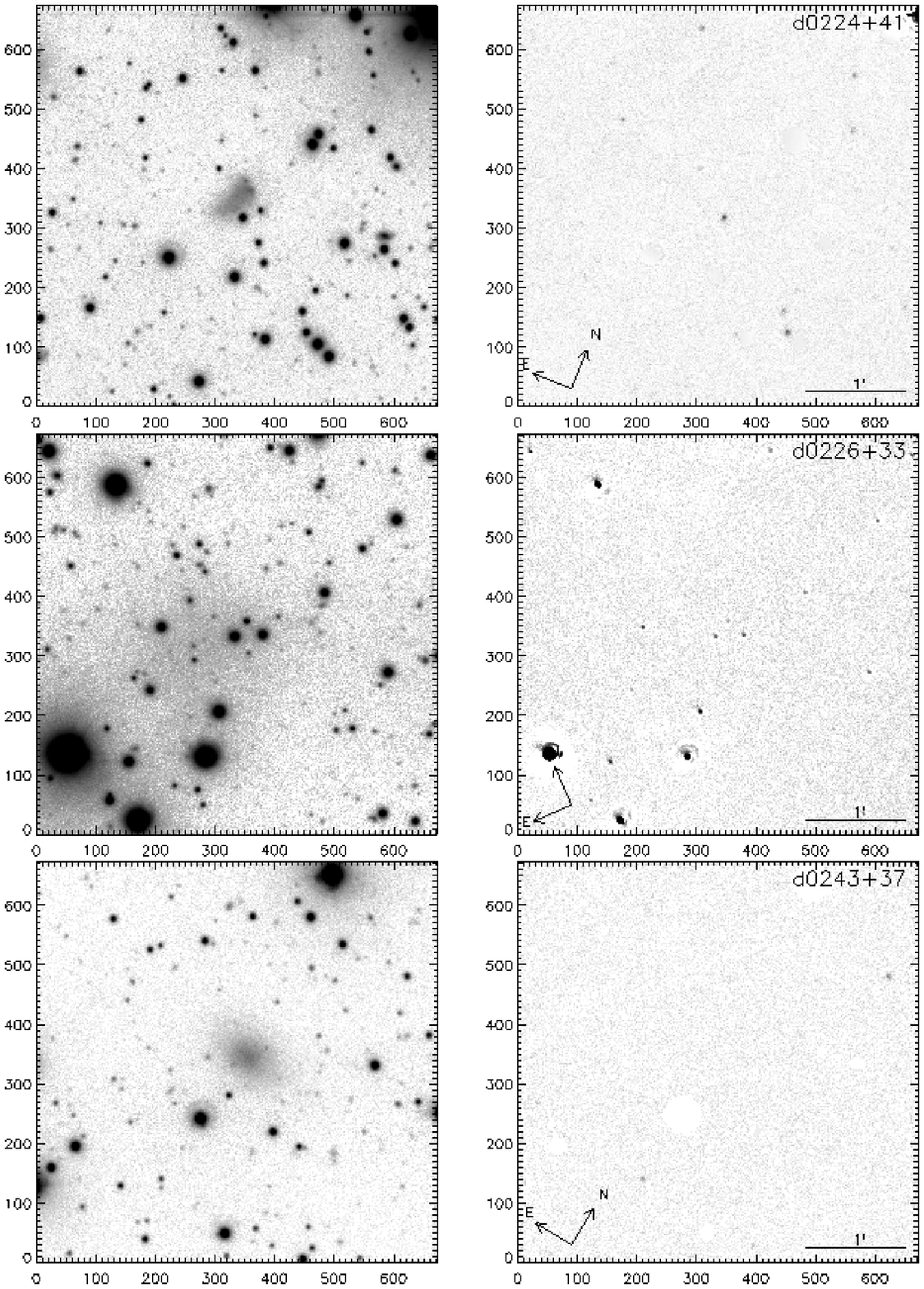}
\end{figure}

\begin{figure}
\includegraphics{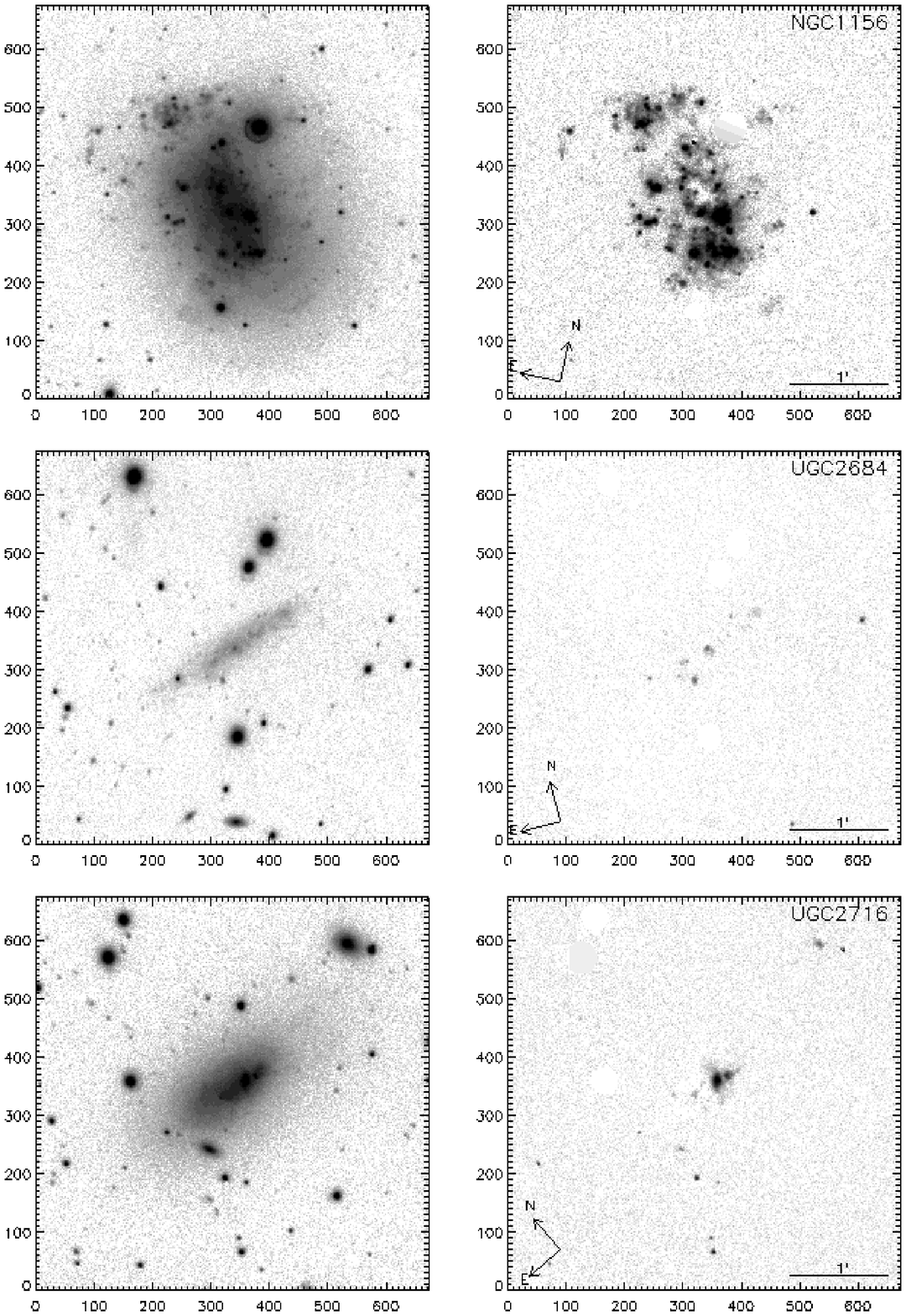}
\end{figure}

\begin{figure}
\includegraphics{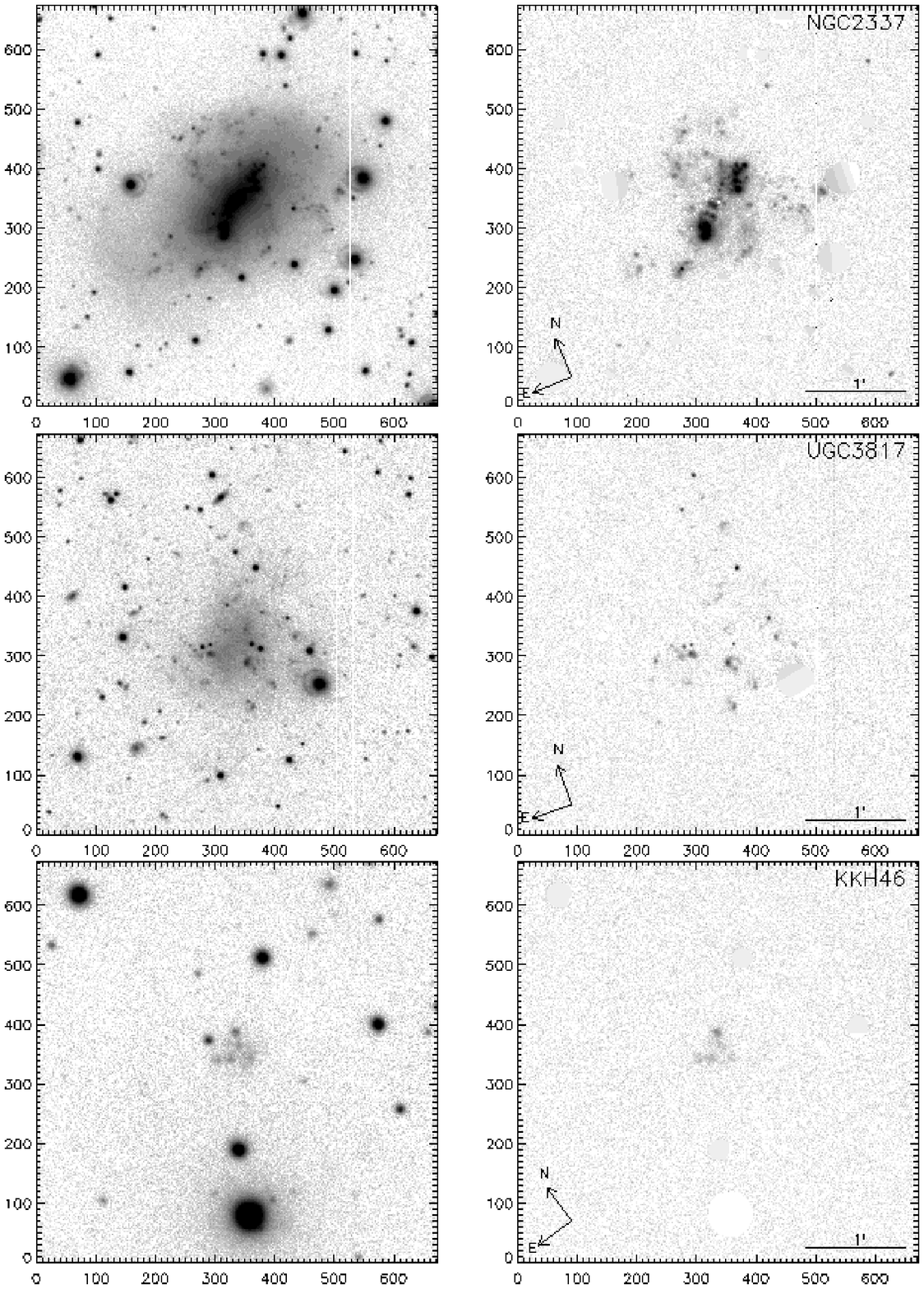}
\end{figure}

\begin{figure}
\includegraphics{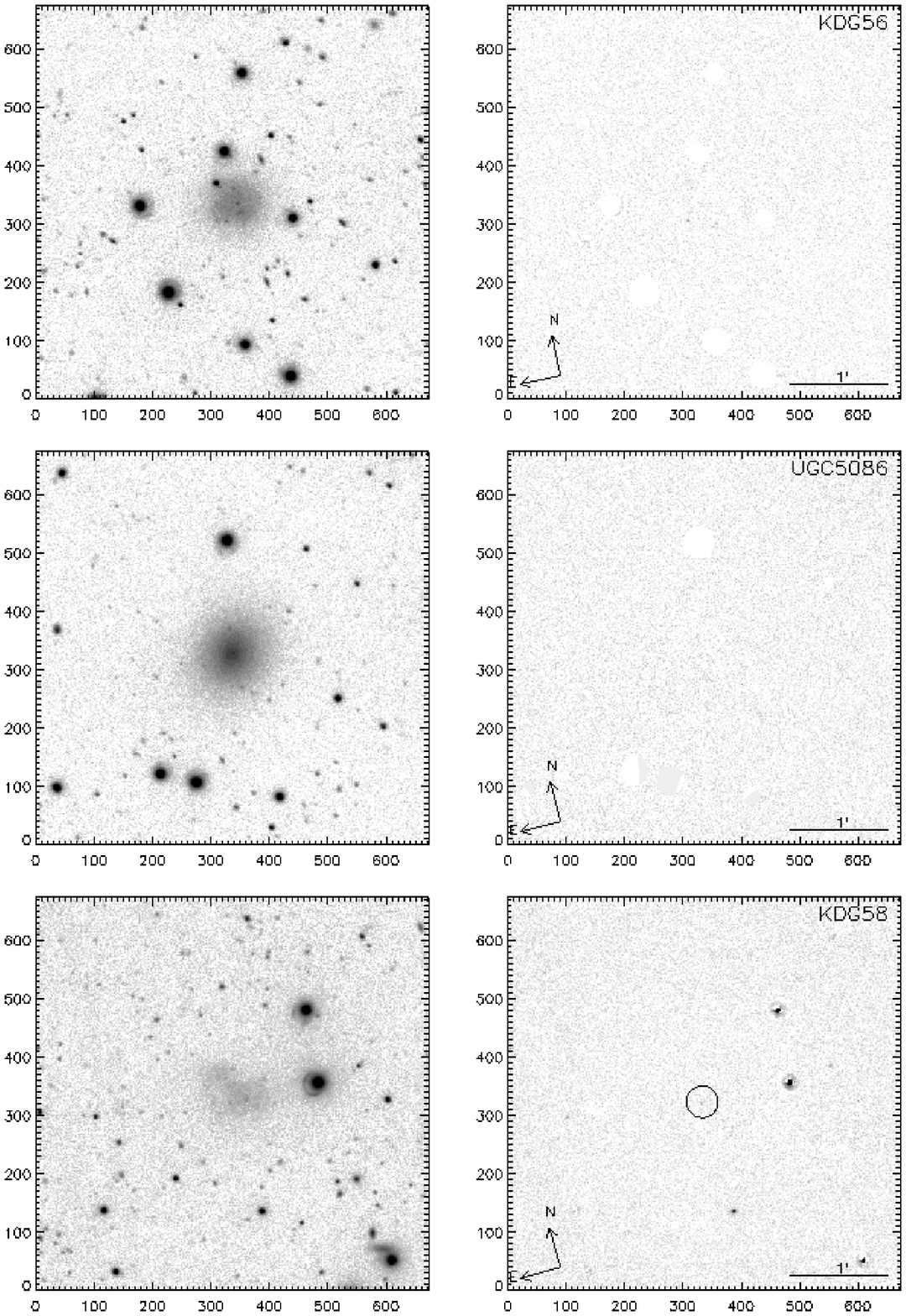}
\end{figure}

\begin{figure}
\includegraphics{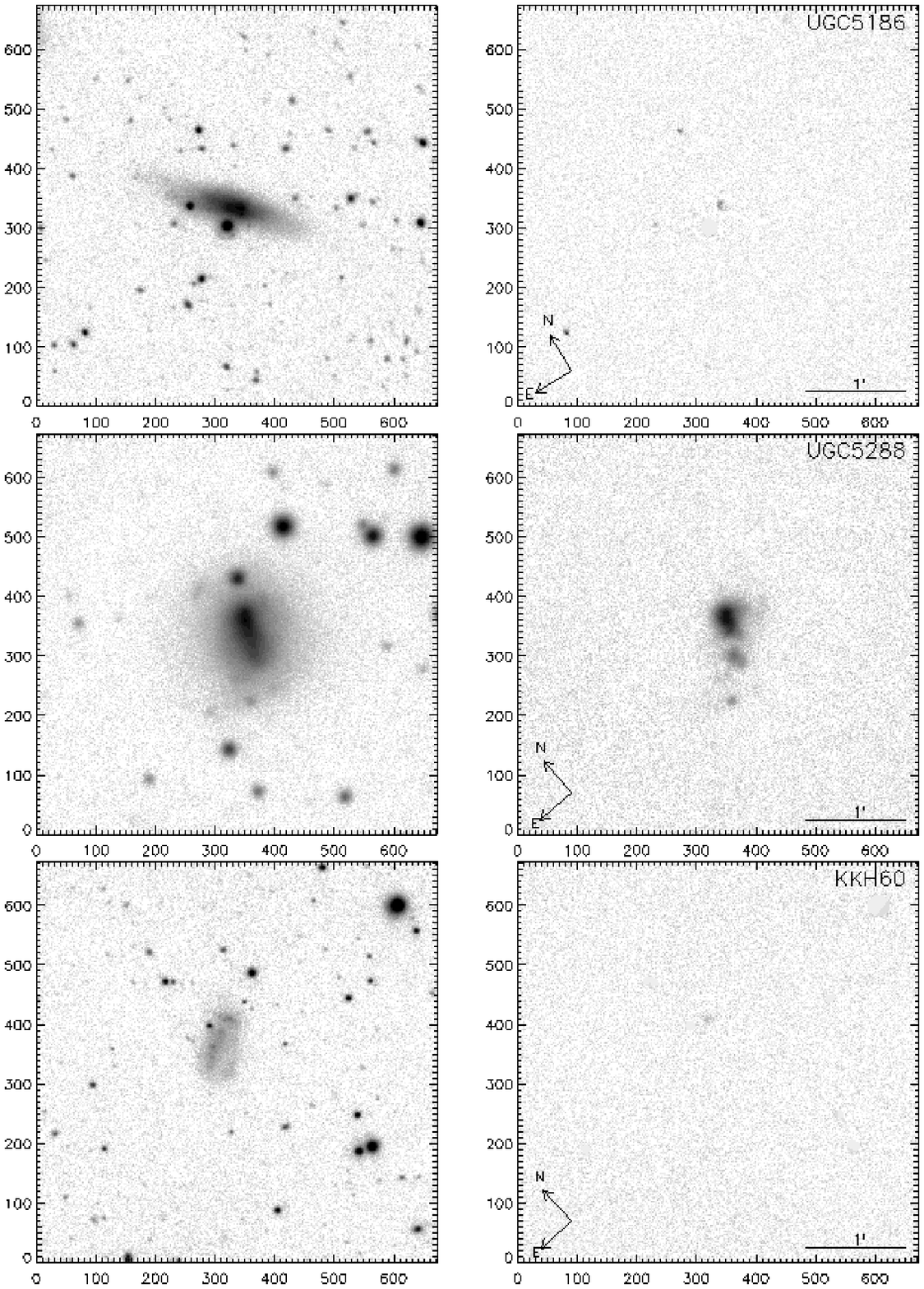}
\end{figure}

\begin{figure}
\includegraphics{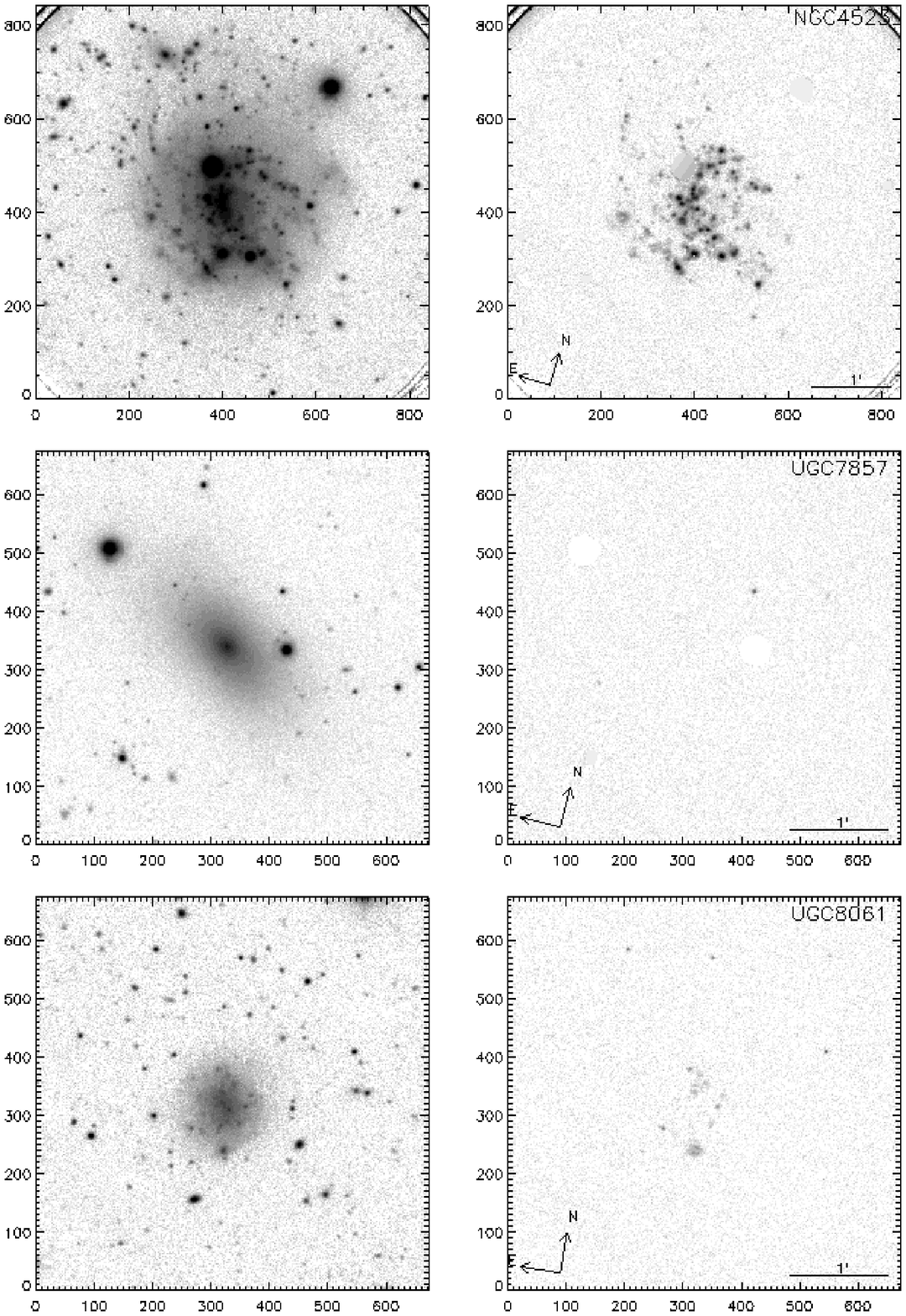}
\end{figure}

\begin{figure}
\includegraphics{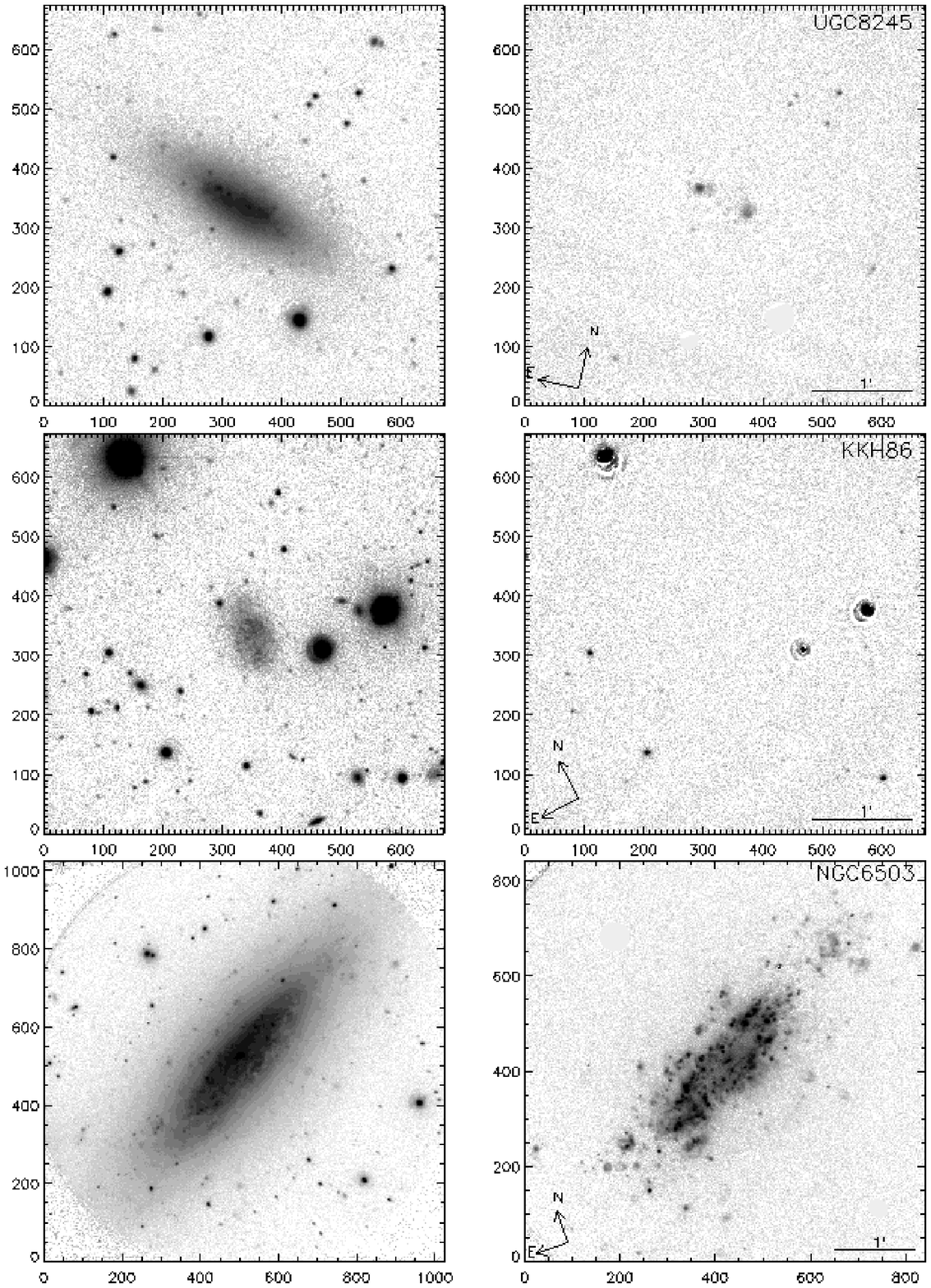}
\end{figure}

\begin{figure}
\includegraphics{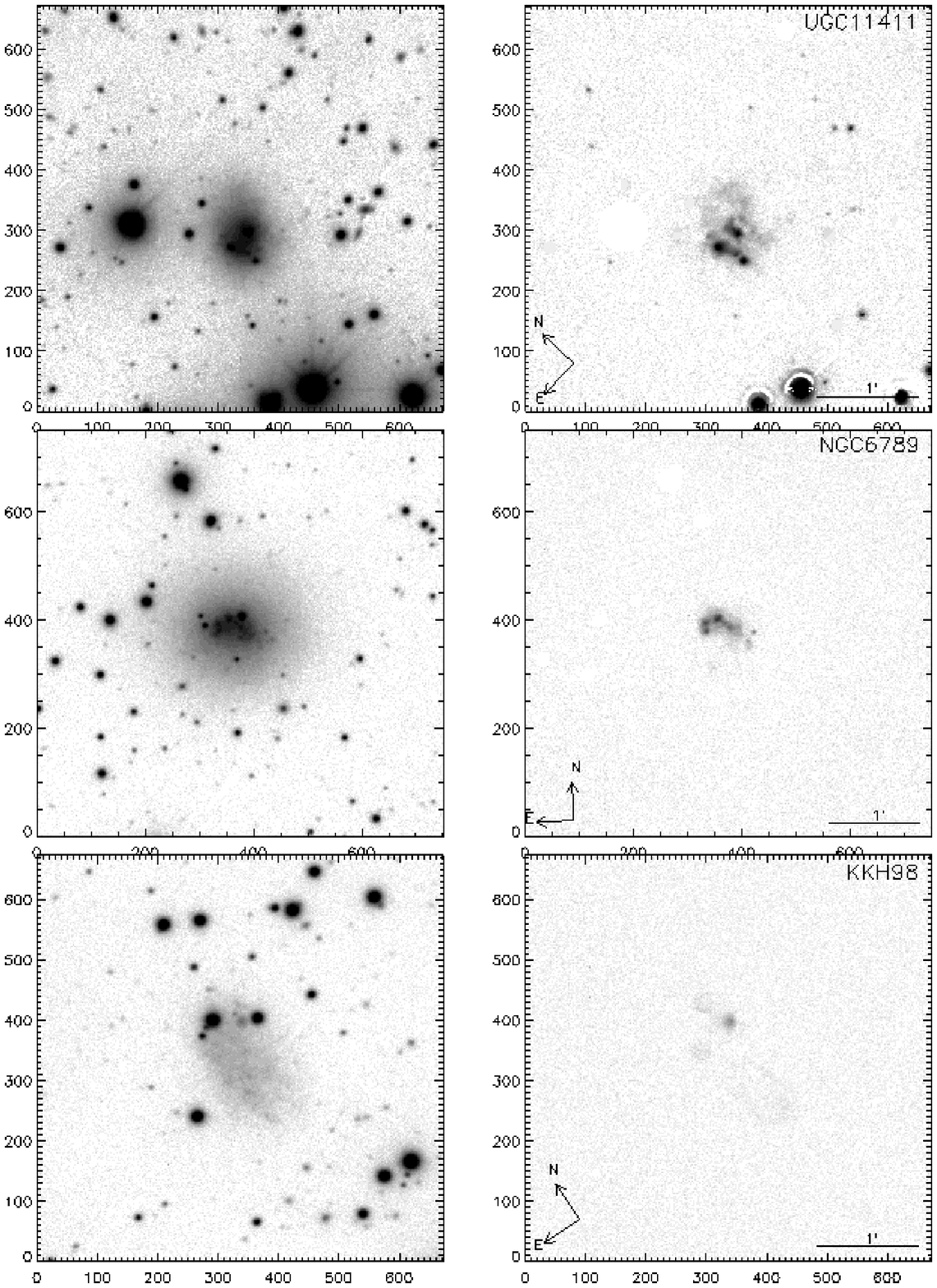}
\end{figure}
\end{document}